\documentclass[pre,twocolumn,showpacs,showkeys,secnumarabic,amssymb,amsmath,unsortedaddress]{revtex4-1}

\usepackage{amssymb,amsmath}
\usepackage{graphicx}
\usepackage{bbold}
\usepackage{bm}
\usepackage{color}

\begin{document}

\title{Non-equilibrium time dynamics of genetic evolution}

\author{Hamid-Reza Rastegar-Sedehi}
\affiliation{State Key Laboratory of Precision Spectroscopy, School of Physical and Material Sciences, East China Normal University, Shanghai 200062, China}
\affiliation{Department of Physics, College of Sciences, Jahrom University, Jahrom, 74137-66171 Iran} 

\author{Chandrashekar Radhakrishnan}
\affiliation{New York University Shanghai, 1555 Century Ave, Pudong, Shanghai 200122, China}

\author{Samer Intissar Nehme}
\affiliation{New York University Shanghai, 1555 Century Ave, Pudong, Shanghai 200122, China}
\affiliation{Department of Chemical Engineering and Biotechnology, University of Cambridge, Pembroke St, Cambridge CB2 3RA, UK} 

\author{Liev Birman}
\affiliation{New York University Shanghai, 1555 Century Ave, Pudong, Shanghai 200122, China}
\affiliation{Department of Physics, University of Wisconsin-Madison, 1150 University Avenue, Madison, Wisconsin 53706, USA}

\author{Paula Velasquez}
\affiliation{New York University Shanghai, 1555 Century Ave, Pudong, Shanghai 200122, China}

\author{Tim Byrnes}
\email{tim.byrnes@nyu.edu}
\affiliation{New York University Shanghai, 1555 Century Ave, Pudong, Shanghai 200122, China}
\affiliation{State Key Laboratory of Precision Spectroscopy, School of Physical and Material Sciences,East China Normal University, Shanghai 200062, China}
\affiliation{NYU-ECNU Institute of Physics at NYU Shanghai, 3663 Zhongshan Road North, Shanghai 200062, China}
\affiliation{National Institute of Informatics, 2-1-2 Hitotsubashi, Chiyoda-ku, Tokyo 101-8430, Japan}
\affiliation{Department of Physics, New York University, New York, NY 10003, USA}

\begin{abstract}
Biological systems are typically highly open, non-equilibrium systems that are very challenging to understand from a statistical mechanics perspective.  
While statistical treatments of evolutionary biological systems have a long and rich history, examination of the time-dependent non-equilibrium dynamics has been less studied.  In this paper we first derive a generalized master equation in the genotype space for diploid organisms incorporating the processes of selection, mutation, recombination, and reproduction.  The master equation is defined in terms of continuous time and can handle an arbitrary number of gene loci and alleles, and can be defined in terms of an absolute population or probabilities.   We examine and analytically solve several prototypical cases which illustrate the interplay of the various processes and discuss the timescales of their evolution.  The entropy production during the evolution towards steady state is calculated and we find that it agrees with predictions from non-equilibrium statistical mechanics where it is large when the population distribution evolves towards a more viable genotype. The stability of the 
non-equilibrium steady state is confirmed using the Glansdorff-Prigogine criterion. 
\end{abstract}

%\pacs{}

\maketitle

\section{Introduction}

The stochastic nature of biological processes can be observed at many levels, for example cell growth \cite{bronk68}, gene expression \cite{volfson05,avlund10}, synaptic plasticity \cite{bienenstock82}, and aging \cite{luciani01}.  The close analogies between statistical mechanics, thermodynamics, and biological systems has yielded many successful studies where ideas from physics can be applied to biological systems \cite{Goel74,brooks88,Web88,Sella05,De11}. In the context of evolutionary biology, there is long history of applying statistics in the field of population genetics \cite{Wri32,Kim83,InSe0502,Gil04}.  For example, the Hardy-Weinberg principle \cite{Hardy1908,GoodRef09Stern43,GoodRef09Weinberg08,Crow99,GoodRef09} predicts the equilibrium population distribution for reproduction of diploid and polyploid organisms.  Stochastic processes such as genetic drift have been analyzed using a  Wright-Fisher model \cite{wright1931evolution,fisher1999genetical,Kim83,Crow71} and 
the neutral theory of molecular evolution \cite{Kim64,InSe0502}.  Following from these pioneering works, numerous studies involving various aspects of genetic evolution have been performed \cite{Gil04,hartl1997principles,crow1970introduction}.  

Recently there has been a renewed interest in relating ideas from non-equilibrium statistical mechanics to biological systems. 
The interest from this point of view has been founded by the seminal results of Jarzynski \cite{jarzynski1997nonequilibrium} and Crooks \cite{crooks1998nonequilibrium,crooks1999entropy} where new quantitative statements could be made relating various quantities to far-from-equilibrium situations.  Studies along this line of research have focused upon a variety of processes including those at the molecular \cite{Nic77,dewey98,qian06,bazzani12,chou2011non}
to the evolutionary level \cite{Doe98,wiley82,brooks88}.  Ideas from non-equilibrium statistical mechanics have shown that they may even help to provide new ways to describe adaptation, self-replication, and other biological processes \cite{england2015dissipative,perunov2016statistical,england2013statistical}. Even for the well-established statistical approaches to evolutionary processes, a further understanding of the time dynamics of these processes is an ongoing area of research \cite{Gin01,KAB14}. In light of the progress that has been made in non-equilibrium statistical mechanics over the last few decades \cite{chou2011non}, a more quantitative approach to understanding non-equilibrium aspects of biological systems continues to be an important topic.  

In this paper, we first derive a master equation in the genotype space which models the evolutionary process taking into account of selection, mutation, recombination, and reproduction. While these have been analyzed in past works involving one or more of these processes, it is desirable to have a compact equation that allows for a generalized analysis \cite{Gil04}.  In particular, our equation can handle an arbitrary number of alleles and gene loci, and is written in a continuous differential form with no ordering of the individual processes.   This is more realistic biologically and is more convenient from a mathematical perspective.  This allows one to handle non-equilibrium situations, and track the time dynamics of evolutionary 
processes in the genotype space.  Other fundamental effects such as gene flow and 
genetic drift are then natural consequences of the model that takes into account the basic process of selection, mutation, and reproduction.  
The Hardy-Weinberg principle is usually defined under a set of restrictions such as 
completely random sexual reproduction, no selection, no mutation, and no genetic drift \cite{Rel2012}.
But in our work we relax these restrictions and explicitly include the effect of these evolutionary 
influences.  Thus our work extends the Hardy-Weinberg principle to a more general context.

Our formulation can be considered as a preliminary step to analyze such population genetics using a non-equilibrium statistical framework.  From the master equation one can calculate the entropy of the distribution, as well as the total entropy production including its environment. We show that the entropy production is consistent with the principles predicted according to recent results in non-equilibrium statistical mechanics, where migration to a more viable population distribution is accompanied by entropy production \cite{perunov2016statistical}.  We show
that the Glansdorff-Prigogine criterion for the stability of non-equilibrium systems is obeyed proving
that the probability distribution achieves a non-equilibrium steady state.

This paper is structured as follows.  In Sec. \ref{sec:model} we introduce the model and our notation for keeping track of the $ M $-loci, $ N $-allele  genotypes.  We then review the fundamental discrete evolutionary processes in Sec. \ref{sec:discreteevolution}.  Based on this we derive the master equation in terms of populations of genotypes in Sec. \ref{sec:populationmaster}, and in terms of relative probabilities in Sec. \ref{sec:probabilitymaster}.  In Sec. \ref{sec:matching} we connect the discrete and continuous formulations in terms of the parameters of the model.  In Sec. \ref{sec:timeevolution} we show numerical tests of our master equation where we verify that it reproduces known results such as the Hardy-Weinberg statistics, selection effects, and mutation.  Here we consider a prototypical viability landscape where there are two highly viable genotypes and discuss the timescale that is required to reach the most viable genotype, and show that interesting dynamics can occur even for such a simplified model. In Sec. \ref{sec:entropy} we discuss the entropy production and stability of the system during the evolution process. Finally,  in Sec. \ref{sec:conc} we summarize our findings.

\begin{figure*}
\includegraphics[width=2.0\columnwidth]{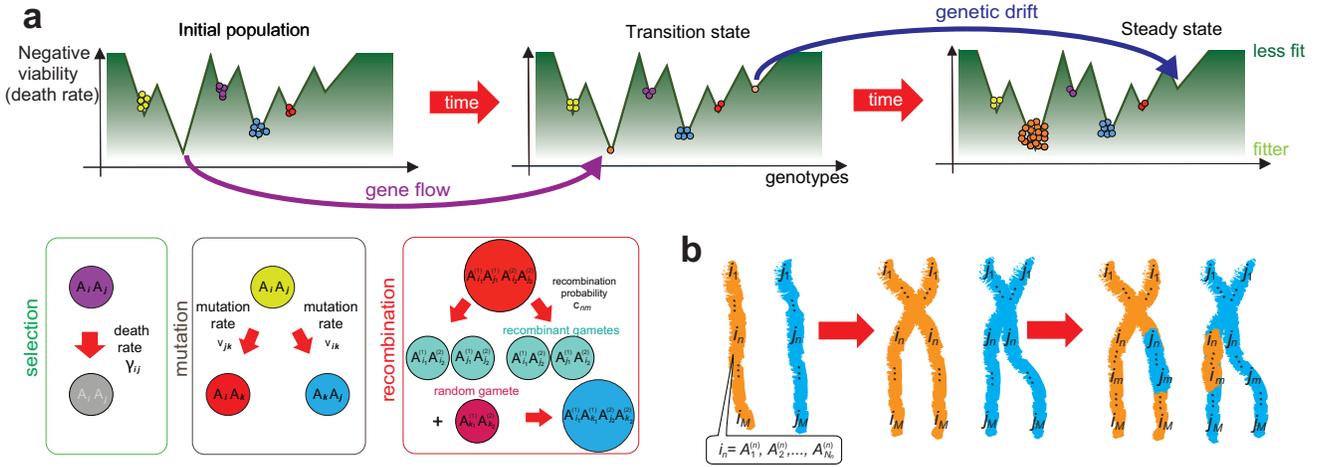}
\caption{\label{fig1} Summary of the genetic evolutionary processes considered in this paper. (a) Processes captured by our master equation (\ref{master}).  Initially various genotypes exist within a fitness landscape, shown here as the negative viability $ - \omega_{ij} $ (equivalent to the death rate).  The population distribution tends to converge towards the fittest genotype, where steady state is reached.  While approaching equilibrium, the processes of selection, mutation, recombination, gene flow, and genetic drift take place. Selection is driven by genotypes with lower viabilities having a higher death rate.  The genotype depicted by the rightmost individual spontaneously appears in the population due to gene flow during the transition state and then drifts off at steady state.  (b) Labeling convention given in this paper depicted on two homologous chromosomes (left), their constituent sister chromatids (center), and after undergoing recombination (right).  The allele configuration on the two chromosomes of a diploid organism is given by (\ref{ijconvention}).  Each chromosome has $M $ loci, with the genotype of the $ n$th locus being labeled by $ i_n, j_n $.  The variables take the values $ i_n, j_n \in 1, \dots N_n $, which correspond to the alleles $ A^{(n)}_{i_n}   A^{(n)}_{j_n} $. }
\end{figure*}

\section{Master equation for genetic evolution}

\subsection{The model}
\label{sec:model}

In this section we set up a model which describes the population and probabilities of a genotype of an $M$-loci, $N$-allele diploid organism \cite{Ewens04}.  The generalization to polyploidy can be performed straightforwardly, but we consider the diploid case for simplicity.   Figure \ref{fig1} shows the global view of our model.  A genotype is specified by the configuration of alleles on two homologous chromosomes according to (see Fig. \ref{fig1}(b))
\begin{align}
\bm{i}& =(i_1,i_2,\dots,i_M) \nonumber \\
\bm{j}& =(j_1,j_2,\dots,j_M) 
\label{ijconvention}
\end{align}
where $ \bm{i} $ represents all the alleles of the different genes on either the maternal or paternal chromosome of one homologous pair, and $ \bm{j} $ encompasses the alleles for its homolog.  The variables $ i_n, j_n $ contain the particular allele for the $n$th gene locus. Denoting the alleles on the $n$th locus, by $ A_{i}^{(n)} A_{j}^{(n)} $, with $ i,j \in 1, \cdots , N_n $, the complete genotype is specified by \cite{Gil04}
\begin{align}
A_{\bm{i}} A_{\bm{j}} \equiv A_{i_1}^{(1)} A_{j_1}^{(1)}
A_{i_2}^{(2)} A_{j_2}^{(2)} \cdots
A_{i_M}^{(M)} A_{j_M}^{(M)} .
\end{align}

The population of the genotype $ A_{\bm{i}} A_{\bm{j}} $ at a given time is denoted
\begin{align}
P_{\bm{i} \bm{j}} \equiv P_{i_1 \dots i_M j_1 \dots j_M} ,
\end{align}
which can take any non-negative number, and is not necessarily an integer.  To give a specific example, this can be the concentration of individuals per unit volume of a sample.  We may normalize the population distribution to define the probability of a genotype $ A_{\bm{i}} A_{\bm{j}} $ occurring according to
\begin{align}
p_{\bm{i} \bm{j}} \equiv p_{i_1 \dots i_M j_1 \dots j_M} = \frac{P_{\bm{i} \bm{j}}}{P_{\text{tot}} } 
\end{align}
where $ P_{\text{tot}} \equiv \sum_{\bm{i} \bm{j}} P_{\bm{i} \bm{j}} $ is the total population. One then has a normalized probability distribution 
\begin{align}
\sum_{\bm{i} \bm{j}} p_{\bm{i} \bm{j}} \equiv 
\sum_{i_1,\dots, i_M} \sum_{j_1,\dots, j_M} p_{i_1 \dots i_M j_1 \dots j_M}  = 1 
\end{align}
and summation runs over  $ i_n,j_n \in 1, \cdots , N_n $.  We assume that the alleles (and their frequencies) of a specific gene are symmetrical by disregarding whether an allele falls on a maternal or paternal chromosome  \cite{Hardy1908,Lewontin58,Gil04,Feldman91}. In other words, $A_{\bm{i}} A_{\bm{j}}$ is equivalent to $ A_{\bm{j}} A_{\bm{i}}$ and $  p_{\bm{i} \bm{j}} = p_{\bm{j}\bm{i} } $. The genes analyzed are thus all autosomal. 

The marginal probabilities on the $ n$th locus is given by summing over the probabilities of the other loci:
\begin{align}
p_{i_n j_n} & = \sum_{\bm{i} \ne i_n \bm{j} \ne i_n  } p_{\bm{i} \bm{j}} \nonumber \\
& \equiv 
\sum_{i_1,\dots, i_{n-1}, i_{n+1},\dots, i_M} \sum_{j_1,\dots, j_{n-1}, j_{n+1},\dots, j_M} p_{i_1 \dots i_M j_1 \dots j_M} .
\end{align}
Thus the probability of a homozygous individual on the $n$th locus is $ p_{i_n j_n} $, while heterozygous individuals have a probability $ p_{i_n j_n}+p_{j_n i_n} $.

\subsection{Discrete evolution}

\label{sec:discreteevolution}

In this section we review the basic evolutionary processes of selection, mutation, and reproduction including recombination. These will be used in the following sections to infer a master equation that captures the evolutionary process in continuous time.

For selection, the probability of the genotypes between generations changes according to \cite{Gil04,Hart07}
\begin{align}
p_{\bm{i} \bm{j}}^{\text{s}} = \frac{\omega_{\bm{i} \bm{j}}}{\bar{\omega}} {p}_{\bm{i} \bm{j}} ,
\label{selectiondiscrete}
\end{align}
where $ p_{\bm{i} \bm{j}}^{\text{s}} $ is the probability of the next generation after selection and $ {\omega}_{\bm{i} \bm{j}} $ is the viability, which is the probability of survival of the individual for genotype $ A_{\bm{i}} A_{\bm{j}} $. The quantity  
\begin{align}
\bar{\omega} = \sum_{\bm{i} \bm{j}} {p}_{\bm{i} \bm{j}} \omega_{\bm{i} \bm{j}} .
\end{align}
is the mean fitness of the population. We note that the viability can be arbitrarily defined by an overall multiplicative constant, hence we typically take it to lie in the range $ 0 \le \omega_{\bm{i} \bm{j}} \le 1 $.

For mutation, the set of alleles $ A_{\bm{i}} $ randomly mutates to $ A_{\bm{i}'} $ with probability $ u_{\bm{i} \bm{i}'} $ \cite{Gil04,Hart07}. Typically the mutations can be point mutations, frame-shift mutations, chromosomal inversions, or translocations as long as they lead to a change in gene expression and are thus not silent. They are also assumed to occur independently on each locus, in which case we may write $ u_{\bm{i} \bm{i}'} = \prod_n u_{i_n i_n'} $. This results in the loss of population from genotype $ A_{\bm{i}} A_{\bm{j}} $ and a gain in the population of $ A_{\bm{i}'} A_{\bm{j}} $.  The former may equally occur for the $ A_{\bm{j}} $ allele, hence one obtains
\begin{align}
p_{\bm{i} \bm{j}}^{\text{m}} = &  {p}_{\bm{i} \bm{j}} -\sum_{ \bm{k} } ({u}_{\bm{j} \bm{k}}+ {u}_{\bm{i} \bm{k}}) {p}_{\bm{i} \bm{j}} + \sum_{\bm{k}} ( {u}_{\bm{k} \bm{j}} {p}_{\bm{i} \bm{k}} +  {u}_{\bm{k} \bm{i}}{p}_{\bm{k} \bm{j}}) , \label{mutationdiscrete}
\end{align}
where $ p_{\bm{i} \bm{j}}^{\text{m}} $ is the new genotype frequency after mutation. The mutation probabilities must be assumed to be sufficiently small such that the coefficient of $ p_{\bm{i} \bm{j}} $ in  (\ref{mutationdiscrete}) is positive, without which we obtain unphysical negative probabilities.

When modeling the reproduction process, we assume that an offspring in the new generation with genotype $A_{\bm{i}} A_{\bm{j}}$ occurs with the product of the probabilities of alleles $A_{\bm{i}} $ and $ A_{\bm{j}} $ in the previous generation \cite{Gil04,Hart07}.  This is
\begin{align}
p_{\bm{i} \bm{j}}^{\text{r}}  = \frac{1}{4} \sum_{\bm{k}} \left( {p}_{\bm{i} \bm{k}}^{\text{c}} + {p}_{\bm{k} \bm{i}}^{\text{c}} \right) \sum_{\bm{l}}  \left( {p}_{\bm{j} \bm{l}}^{\text{c}} +{p}_{\bm{l} \bm{j}}^{\text{c}}  \right)   ,
\label{reproductiondiscrete}
\end{align}
where $ p_{\bm{i} \bm{j}}^{\text{r}}  $ is the new allele frequency after reproduction. This is the same principle that leads to Hardy-Weinberg statistics, which we verify in Sec.  \ref{sec:timeevolution}.  

The $ {p}_{\bm{i} \bm{k}}^{\text{c}} $ on the right hand side of (\ref{reproductiondiscrete}) takes into account the effects of recombination that can potentially occur during meiosis thereby increasing the genetic diversity in a population over time by allowing for Mendelian independent assortment of genes \cite{Gei44}.  However, if the loci of two or more genes fall very close to one another, they are said to be linked and Mendelian independent assortment no longer applies. When modeling recombination, we take into account both independent assortment and gene linkage by considering a chromosome segment between the $n$th and $m$th locus interchanging:
\begin{align}
A_{\bm{i}} A_{\bm{j}} \rightarrow A_{\bm{I}_{nm} } A_{\bm{J}_{nm} }
\end{align}
where the recombinant functions are
\begin{align}
\bm{I}_{nm}  \equiv (i_1, \dots, i_{n-1}, j_{n}, \dots, j_{m},i_{m+1},\dots, i_M) \nonumber \\
\bm{J}_{nm} \equiv (j_1, \dots, j_{n-1}, i_{n}, \dots, i_{m},j_{m+1},\dots, j_M)
\end{align}
as also illustrated in Fig. \ref{fig1}(b).  The probability evolves in a similar way to mutation, and takes the form 
\begin{align}
p_{\bm{i} \bm{j}}^{\text{c}}= p_{\bm{i} \bm{j}} 
- \sum_{n,m=1}^M c_{nm} p_{\bm{i} \bm{j}} + \sum_{n,m=1}^M c_{nm} p_{\bm{I}_{nm}  \bm{J}_{nm} },
\label{recombprob}
\end{align}
where $  c_{nm} $ is the probability that the recombination occurs between the $n$th and $m$th loci. We consider recombination only within (\ref{reproductiondiscrete}) since it only contributes to genetic diversity during the process of meiosis.  

In the discrete approach, one cycle of selection, mutation, and reproduction including recombination is then calculated by setting the output of (\ref{selectiondiscrete}) as the input of (\ref{mutationdiscrete}), then in turn setting this as the input of (\ref{reproductiondiscrete}). This process can be repeated until a steady-state probability distribution is reached.

\subsection{Continuous time: Population master equation}

\label{sec:populationmaster}

We now deduce a continuous time form of  (\ref{selectiondiscrete}), (\ref{mutationdiscrete}), and (\ref{reproductiondiscrete}) to obtain a master equation which describes the selection, mutation, and the reproductive process including recombination.  A continuous time form is more realistic because all these processes occur simultaneously in a biological system and not in discrete steps as implied by the above equations.  More importantly, this form makes analysis using the techniques of non-equilibrium statistical mechanics easier and the behavior of the model more apparent. 

By examining (\ref{selectiondiscrete}), (\ref{mutationdiscrete}), and (\ref{reproductiondiscrete}) we deduce that the population  distribution should evolve as
\begin{align}
\frac{dP_{\bm{i} \bm{j}}}{dt} = & - \gamma_{\bm{i} \bm{j}} P_{\bm{i} \bm{j}} - \gamma'  P_{\text{tot}} P_{\bm{i} \bm{j}} \nonumber \\
& - \sum_{\bm{k}} ({v}_{\bm{j} \bm{k}}+ {v}_{\bm{i} \bm{k}})  P_{\bm{i} \bm{j}} + \sum_{\bm{k}} ( v_{\bm{k} \bm{j}} P_{\bm{i} \bm{k}} +  v_{\bm{k} \bm{i}} P_{\bm{k} \bm{j}}) \nonumber \\
& + \frac{r}{4P_{\text{tot}} } \sum_{\bm{k}} \left( P_{\bm{i} \bm{k}}^{\text{c}}+P_{\bm{k} \bm{i}}^{\text{c}} \right) \sum_{\bm{l}} \left( P_{\bm{j} \bm{l}}^{\text{c}} + P_{\bm{l} \bm{j}}^{\text{c}} \right) .
\label{populationmaster}
\end{align}
Here $ \gamma_{\bm{i} \bm{j}} $ is the death rate of genotype $ A_{\bm{i}} A_{\bm{j}} $, and we show below is an equivalent way of writing the selection process (\ref{selectiondiscrete}).  There is no factor of $ \bar{\omega} $ as in (\ref{selectiondiscrete}), as there is no need to normalize a population distribution.  The recombinant populations are given by
\begin{align}
P_{\bm{i} \bm{j}}^{\text{c}}= P_{\bm{i} \bm{j}} 
- \sum_{n,m=1}^M c_{nm} P_{\bm{i} \bm{j}} + \sum_{n,m=1}^M c_{nm} P_{\bm{I}_{nm}  \bm{J}_{nm} }. 
\end{align}
The terms in the second line of (\ref{populationmaster}) are the equivalent of (\ref{mutationdiscrete}), except that the $ {v}_{\bm{i} \bm{k}} $ are mutation rates - the rates at which new mutations appear in a population.  The last line of  (\ref{populationmaster}) has the same form as (\ref{reproductiondiscrete}), except that we add a reproduction rate $ r $.  Finally, the extra term $  \gamma' $ is the additional death rate per total population which sets an upper limit to the population growth. Without this term the population either grows exponentially without bound, or decays to zero.  In realistic systems there is such a decay term as there is a limit to the resources that sustain a population (i.e. overpopulation effects).

The population master equation (\ref{populationmaster}) is potentially useful in situations where one would like to deal in terms of actual numbers of individuals, rather than probabilities.  At steady-state, we have $ \frac{dP_{\bm{i} \bm{j}}}{dt} = 0 $, and we can sum over all $ \bm{i}, \bm{j} $ in  (\ref{populationmaster}) to obtain
\begin{align}
P_{\text{tot}}  = \frac{r - \bar{\gamma}}{\gamma'} ,
\end{align}
where $ \bar{\gamma} \equiv \sum_{\bm{i} \bm{j}} \gamma_{\bm{i} \bm{j}} p_{\bm{i} \bm{j}} $ is the average death rate. As the total population must be positive, we have $ r > \bar{\gamma} $ which states that the reproductive rate must be larger than the average death rate.  If this is not satisfied, the total population converges to $ P_{\text{tot}}  = 0 $, (i.e. extinction). 

We note that the meaning of (\ref{populationmaster}) is in terms of an average over many stochastic instances during the time dynamics.  For example, exponential growth of a finite number of individuals in a single run of the experiment increases in a stochastic fashion, only giving the smooth exponential behavior after averaging over many runs.  Put another way,  (\ref{populationmaster}) does not contain any fluctuations in the population numbers, but does take into account of fluctuations in the genotype.  To recover the dynamics dictated by (\ref{populationmaster}) from experimental data, one must average over many stochastic instances under the same parameters and compare the distribution at each point in time.  This is equally true of the probabilitistic master equation that will be derived below.

\subsection{Continuous time: Probabilistic master equation}

\label{sec:probabilitymaster}

We now require an equivalent equation to (\ref{populationmaster}) for the probability $ p_{\bm{i} \bm{j}} $.  The main requirement is that unlike the population master equation, the probability equation must preserve normalization $ \sum_{\bm{i} \bm{j}} p_{\bm{i} \bm{j}} = 1 $ throughout the time evolution. For mutation, since for each loss there is a corresponding gain, thus detailed balance is obeyed and probability is conserved.  However, the death and reproduction terms do not and thus detailed balance is not obeyed.  This can be remedied by adding normalization terms.  According to the definition $ p_{ij} = P_{ij}/P_{\text{tot}} $, we have
\begin{align}
\frac{d p_{ij}}{dt} = \frac{1}{P_{\text{tot}}} \frac{d P_{ij}}{dt} - \frac{p_{ij}}{P_{\text{tot}}} \frac{d P_{\text{tot}}}{dt}
\label{probderive}
\end{align}
since both $ P_{ij} $ and $P_{\text{tot}} $ change with time.  To obtain $ \frac{d P_{\text{tot}}}{dt} $, we take a sum of (\ref{populationmaster}) over $ i,j $ to obtain
\begin{align}
 \frac{d P_{\text{tot}}}{dt} = - \bar{\gamma} P_{\text{tot}}  - \gamma' P_{\text{tot}}^2 + r P_{\text{tot}}
\label{popderiv}
\end{align}
where all the mutation terms cancel due to detailed balance. Substitution of (\ref{populationmaster}) and (\ref{popderiv}) into (\ref{probderive}) gives the probabilistic master equation
\begin{align}
\frac{dp_{\bm{i} \bm{j}}}{dt} = & - \gamma_{\bm{i} \bm{j}} p_{\bm{i} \bm{j}} + \bar{\gamma}  p_{\bm{i} \bm{j}} \nonumber \\
& - \sum_{\bm{k}} ({v}_{\bm{j} \bm{k}}+ {v}_{\bm{i} \bm{k}}) p_{\bm{i} \bm{j}} + \sum_{\bm{k}} ( v_{\bm{k} \bm{j}} p_{\bm{i} \bm{k}} +  v_{\bm{k} \bm{i}} p_{\bm{k} \bm{j}}) \nonumber \\
& + \frac{r}{4} \sum_{\bm{k}} \left( p_{\bm{i} \bm{k}}^{\text{c}} +p_{\bm{k} \bm{i}}^{\text{c}} \right) \sum_{\bm{l}}\left( p_{\bm{j} \bm{l}}^{\text{c}} + p_{\bm{l} \bm{j}}^{\text{c}}\right) - r p_{\bm{i} \bm{j}},
\label{master}
\end{align}
where the recombinant probabilities are given by (\ref{recombprob}). 

The main difference of the above to (\ref{populationmaster}) is the presence of two extra terms proportional to $ r $ and $ \bar{\gamma} $ which play the role of keeping the probability distribution $ p_{\bm{i} \bm{j}} $ normalized.  We also note that the nonlinear death rate $ \gamma' $ plays no role in the probability equation as it is canceled by the normalizing terms.  The master equation (\ref{master}) is evolved until a steady-state is reached.  While in principle it is possible that the form of the master equation (\ref{master}) to not have a steady-state,  for biologically relevant parameters where mutation rates are less than the reproduction and death rates, we find that steady-state is attained for sufficiently long propagation times.

\subsection{Matching discrete evolution to continuous evolution}
\label{sec:matching}

The master equation (\ref{master}) has a different set of parameters to those in the discrete versions (\ref{selectiondiscrete}), (\ref{mutationdiscrete}), and (\ref{reproductiondiscrete}).  In this section we show the relationship between these. 
 We examine the selection, mutation, and recombination formulas by considering the relevant terms in (\ref{master}) separately.  Taking the time between generations in the discrete formulation as $ \Delta t $, the new probability distribution under selection is
\begin{align}
p_{ij}^{\text{s}} \approx p_{ij} + \Delta t \frac{d p_{ij}}{dt} & = ( 1 -  \Delta t \gamma_{ij} + \Delta t \bar{\gamma} ) p_{ij} \nonumber \\
& \approx \frac{ ( 1-  \Delta t \gamma_{ij}) p_{ij}}{1-  \Delta t \bar{\gamma}} ,
\end{align}
where in the last line a Taylor expansion is performed on the denominator, assuming that  $ \Delta t \bar{\gamma} \ll 1  $.  Comparing this to (\ref{selectiondiscrete}), we obtain $ \omega_{ij} = 1- \Delta t \gamma_{ij} $. This equation indicates that the survival probability $ \omega_{ij} $ is equal to one minus the death probability $ \gamma_{ij} \Delta t $.  

Similarly, for mutation, we have
\begin{align}
p_{ij}^{\text{m}} & \approx p_{ij} + \Delta t \frac{d p_{ij}}{dt} \nonumber \\
& =  {p}_{ij} - \Delta t \sum_{k=1}^{N} ({v}_{jk}+ {v}_{ik}) {p}_{ij} + \Delta t \sum_{k=1}^{N} ( {v}_{kj}{p}_{ik} +  {v}_{ki}{p}_{kj}). 
\end{align}
On comparison with (\ref{mutationdiscrete}) we have $ u_{ij} =  v_{ij} \Delta t $.  

For recombination, we have 
\begin{align}
p_{ij}^{\text{r}} & \approx p_{ij} + \Delta t \frac{d p_{ij}}{dt}  \nonumber \\
& =  (1- r \Delta t) {p}_{ij} + \frac{r \Delta t}{4} \sum_{k=1}^{N} \left( {p}_{ik}+ {p}_{ki}\right) \sum_{k=1}^{N} \left( {p}_{jk}+{p}_{kj}\right) .  
\end{align}
In the discrete case (\ref{reproductiondiscrete}), it is assumed that the new generation entirely replaces the previous generation, which occurs at $ r \Delta t = 1 $.  We can also view this as the timescale of the discrete evolution being set by the reproductive rate.  

In summary, the parameters in the discrete and continuous evolution can be related by
\begin{align}
\omega_{\bm{i} \bm{j}} & = 1 - \gamma_{\bm{i} \bm{j}} \Delta t, \label{omegagamma} \\
u_{\bm{i} \bm{j}} & = v_{\bm{i} \bm{j}} \Delta t \label{uv} \\
r \Delta t & = 1  \label{rdeltat},
\end{align}
where $ \Delta t $ is the time between generations. We note that the discrete and continuous evolution will in general give 
different time dynamics.  They will only coincide under certain assumptions as outlined above, and therefore the equivalence (\ref{rdeltat}) is only in this context. The effect of each of these parameters are summarized in Fig. \ref{fig1}(a).

\begin{figure}
\includegraphics[width=\columnwidth]{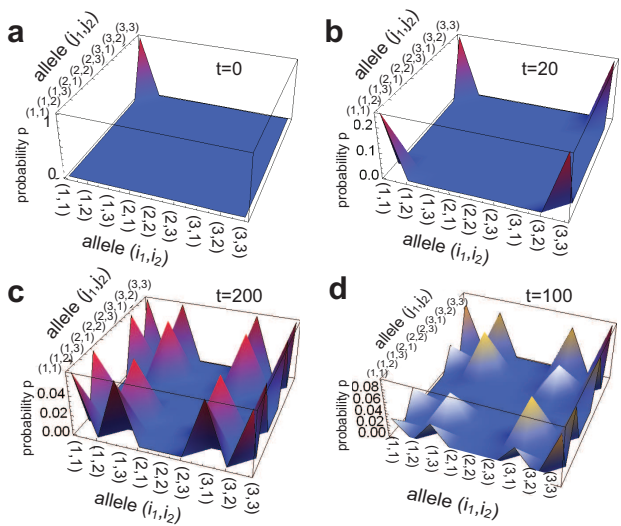}
\caption{	\label{fig2} Time evolution of the master equation for evolution of $ M = 2$ gene loci under various conditions.  (a)(b)(c) Evolution under random mating ($r= 1 $) with no selection or mutation ($ \gamma_{ij} = v_{ij} = 0 $) for two loci $ M = 2 $ and three allele variations $ N =3 $ with a linear viability function. (a)The initial condition where the entire population has the genotype $ A_1^{(1)} A_3^{(1)} A_1^{(2)} A_3^{(2)} $.  (b) Steady-state probability ($ t = 20 $) with no recombination $ c_{nm}= 0 $.  (c)  Steady-state probability ($ t = 200 $) with recombination $ c_{11}= 1 $, otherwise $c_{nm} =0$. (d) Steady-state probability ($ t = 100 $) with recombination $ c_{11}= 1 $ and selection  ($ \gamma_{ij} =  1 $) for two loci $ M = 2 $ and three allele variations $ N =3 $ with the same viability function defined in (\ref{viability}) and parameters a = - 0.5, b = 0.1, c = 0.7.  }
\end{figure}

\section{Time evolution}

\label{sec:timeevolution}

\subsection{Hardy-Weinberg statistics}

We now show some basic properties of the master equation (\ref{master}).  We first examine the effect of the reproduction and recombination terms under Hardy-Weinberg assumptions of no selection ($\gamma_{ij} = 0 $), mutation ($ v_{ik} = 0 $), gene flow, or genetic drift. We will examine a two locus example ($ M = 2 $).  In this case the allele labels involve two variables and we have $ \bm{i} = (i_1, i_2),  \bm{j}   = (j_1, j_2)$. To verify Hardy-Weinberg statistics, we evolve an initially completely heterozygous distribution with all the population in the genotype $ A_1^{(1)} A_3^{(1)} A_1^{(2)} A_3^{(2)} $ with no recombination.  Fig. \ref{fig2}(a)(b) shows the evolution towards equilibrium, which eventually converges towards Hardy-Weinberg statistics of 
\begin{align}
p_{(1,1)(1,1)} = p_{(1,3)(1,3)}= p_{(3,1)(3,1)}= p_{(3,3)(3,3)} =1/4,
\end{align}
corresponding to an equal population of 
\begin{align}
& A_1^{(1)} A_1^{(1)} A_1^{(2)} A_1^{(2)},  A_1^{(1)} A_3^{(1)} A_1^{(2)} A_3^{(2)}, \nonumber \\
& A_3^{(1)} A_1^{(1)} A_3^{(2)} A_1^{(2)}, A_3^{(1)} A_3^{(1)} A_3^{(2)} A_3^{(2)} .
\end{align}
More generally, we find the effect of the reproductive term is to evolve the probability towards
\begin{align}
p_{\bm{i}\bm{j}} (t \rightarrow \infty) & = \rho_{\bm{i}}  (t=0)  \rho_{\bm{i}} (t=0) \label{steadystateprob} 
\end{align}
where $ \rho_{\bm{i}}(t) = \frac{1}{2} \sum_{\bm{j}} ( p_{\bm{i}\bm{j}} (t ) + p_{\bm{j}\bm{i}} (t ) ) $ is the probability of allele $ A_{\bm{i}} $.  The substitution of (\ref{steadystateprob}) into the reproductive terms of (\ref{master}) shows that these terms cancel giving $ \frac{d p_{\bm{i}\bm{j}}}{dt} = 0 $.  This indicates that the reproductive terms comply with the Hardy-Weinberg principle. 

Including recombination creates more diversity as illustrated in Fig. \ref{fig2}(c). Starting from the same initial condition as Fig. \ref{fig2}(a)  gives a steady state distribution where each allele on each locus can be either $ A_1 $ or $ A_3 $ thereby giving equal probabilities for all $ 4^2=16 $ combinations of  $ A_{1,3}^{(1)} A_{1,3}^{(1)} A_{1,3}^{(2)} A_{1,3}^{(2)}$. As expected, recombination thus results in an increase of genetic diversity, in agreement with past studies on related models 
\cite{Per79,Per77,Shp2002}.

\subsection{Selection between competing genotypes}

\begin{figure}
\includegraphics[width=\columnwidth]{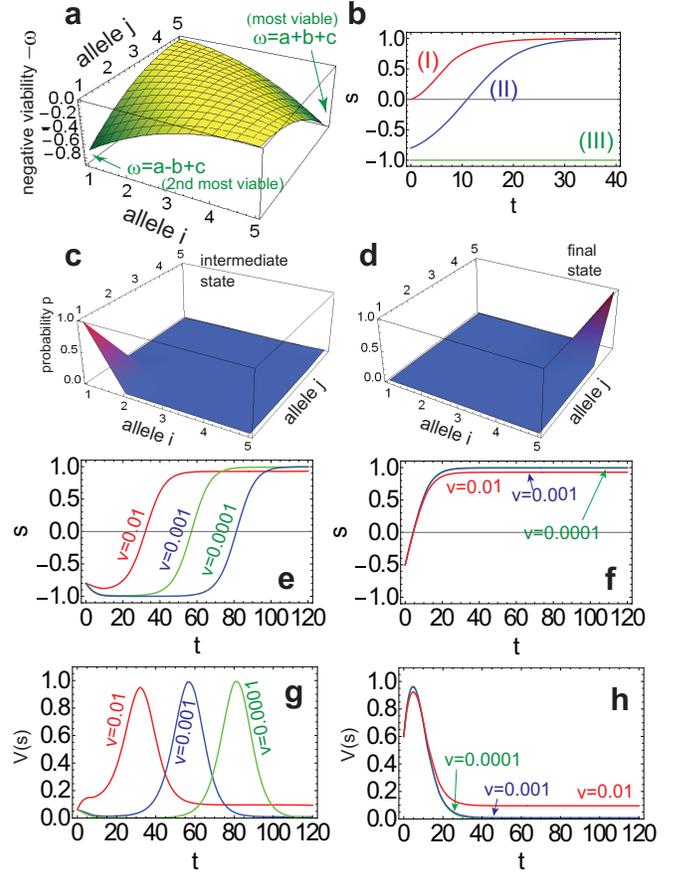}
\caption{	\label{fig3} The effects of different parameters on a fitness landscape of one gene locus with two genotypes with high viability.(a)(b) Evolution for pure selection for one locus $ M = 1 $ and five allele variations $ N = 5 $.  (a) Viability distribution for simulations in Fig. \ref{fig3} with parameters $ a = 0.8 $, $ b= 0.1 $, $ c= 0 $. (b) Time evolution of the variable $ s=s_{\text{m}} = s_{\text{f}} $ from different initial conditions: (I) $ p_{ij}(t=0) = 1/N^2 $; (II) $ p_{ij}(t=0) = 0.9 \delta_{i,1} \delta_{j,1} + 0.1 \delta_{i,N} \delta_{j,N} $; (III)  $ p_{ij}(t=0) = \delta_{i,1} \delta_{j,1} $. Here $ \delta_{i,j} $ is the Kronecker delta. (c)(d)(e)(f)(g)(h) Evolution for mutation for one locus $ M = 1 $ and five allele variations $ N = 5 $. (c)(d) Probability distribution at times (c)  $ t = 60 $ and (d) $ t = 120 $ starting from an initial configuration $ p_{ij}(t=0) = 0.4 \delta_{i,1} \delta_{j,1} + 0.2 (\delta_{i,1} \delta_{j,2} + \delta_{i,2} \delta_{j,1} + \delta_{i,2} \delta_{j,2}) $. Identically, (d) is also the landscape obtained at $ t = 20 $ for the time evolution under pure selection ($\gamma_{ij} = 1 - \omega_{ij} $) and no mutation or reproduction ($ v_{ij} = r= 0 $) with an  initial condition $ p_{ij}(t=0) = 1/N^2 $. 
After a time $ t \sim 10 $ the distribution becomes entirely dispersed at $ i = j = 1 $, until a time $ t =  t_{\text{c}} \sim 70 $ (where $ t_{\text{c}} $ is the critical wait time), when the population of the most viable genotype $ i = j = 5 $ starts to grow. After a time $ t_{\text{c}} $, the most viable genotype becomes stable. The parameters used are $ N= 5 $, $ a = 0.9 $, 
$ b = 0.1 $, $ c = 0 $, $ r = 0.01 $, $ v_{ij} = v ( 1 - \delta_{ij}) $, $ v = 10^{-4} $. (e) The average value 
$ s = \bar{s}_{\text{m}} = \bar{s}_{\text{f}} $ for the same parameters as above but with the mutation rates 
$ u $ as marked. (f) is the same as (e) but with the initial condition $ p_{ij}(t=0) = 0.4 \delta_{i,1} \delta_{j,1} + 0.2 (\delta_{i,1} \delta_{j,2} + \delta_{i,2} \delta_{j,1} + \delta_{i,5} \delta_{j,5}) $, illustrating gene flow.(g) shows the variances $ V(s) = V(s_{\text m}) = V(s_{\text f}) $ for the same parameters as above but with the mutation rates $ u $ as marked. (h) is the same as (g) but with the initial condition given in (f). }
\end{figure}

Now let us turn to how the master equation behaves under selection. For the remainder of this section we consider single locus $ M = 1 $ case for simplicity.  Our allele indices therefore single variables $ \bm{i} = i, \bm{j} = j $.  A prototypical form of the viability is assumed
\begin{align}
{\omega}_{ij} =  a (s_{\text{m}} +s_{\text{f}})^{2} + b (s_{\text{m}}+s_{\text{f}}) + c ,
\label{viability}
\end{align}
where $ a,b,c $ are constants and 
\begin{align}
 s_{\text{m}} & \equiv \frac{2i-N-1}{N-1} \nonumber \\
 s_{\text{f}} & \equiv \frac{2j-N-1}{N-1}
\end{align}
are variables that we define for convenience that identify the allele types within a range $ s_{\text{m,f}} \in [-1,1] $. 
For the example shown in Fig. \ref{fig3}(a), the form of the viability (\ref{viability}) is taken such that there are two genotypes of high viability, i.e. low death rates for $A_1 A_1 $ and $A_N A_N $.  

In Fig. \ref{fig3}(b)(c)(d) we examine the effect of only selection in (\ref{master}) with two homozygous genotypes $ A_1 A_1 $ (or $  s_{\text{m}} = s_{\text{f}} = -1 $) and $ A_N A_N $ (or $  s_{\text{m}} = s_{\text{f}} = 1 $)  having a high viability. As one would expect the population becomes generally more distributed to the more viable genotype, and reaches steady state for long evolution times.  Fig. \ref{fig2}(d) shows another example where selection effects together with recombination.  In comparison with the case without selection Fig. \ref{fig2}(c), the population distribution reaches equilibrium with a bias towards the most viable genotypes. 

The dynamics tend to evolve on a timescale $\tau \propto 1/b $.  As indicated in Fig. \ref{fig3}(a), $ b $ is the parameter that determines the difference in viability between the two dominant genotypes.  Thus we can deduce that one of the timescales of the evolutionary process is determined by
\begin{align}
\tau \sim \frac{1}{\Delta \gamma}
\label{timeselection}
\end{align}
where $ \Delta \gamma $ is the difference in death rate between the two dominant genotypes, and in the case of Fig. \ref{fig3}(a) is equal to $ \Delta \gamma = \gamma_{11} - \gamma_{NN}$. This is based on the assumption that mutations leading to a higher viability in a specific subgroup of a population will cause a decrease in their overall death rate. However, depending on the initial conditions -- such as those shown in Fig. \ref{fig3}(b) -- this may not always be true.  We also repeat the time evolution for the case where the entire population starts at $ A_1 A_1 $ - the less fitter genotype of the two (case III).  In this case the population distribution is completely static. It does not evolve towards the fitter $ A_N A_N $ genotype due to the lack of genetic diversity in the original distribution.  

The above behavior can be understood by obtaining an equation for the 
dynamics of the  $ s_{\text{m}},s_{\text{f}} $ variables. Multiplying (\ref{master}) by $ s_{\text{m}} $ and summing over $ i,j $ we obtain an equation
\begin{align}
\frac{d \bar{s}_{\text{m}} }{dt}  = &\frac{a}{4} \left[ C(s_{\text{m}}^2,s_{\text{m}}) + C(s_{\text{m}}s_{\text{f}},s_{\text{m}}) + C(s_{\text{f}}^2,s_{\text{m}}) \right]  \nonumber \\
& + \frac{b}{2} \left[ V ( s_{\text{m}}) + C(s_{\text{f}},s_{\text{m}}) \right] 
\label{selectioneffective0} \\
\approx &  [b+a( \bar{s}_{\text{m}} + \bar{s}_{\text{f}} ) ] V ( s_{\text{m}})/2
\label{selectioneffective}
\end{align}
where $ \bar{s}_{\text{m}} $ is the average of $ s_{\text{m}} $, $ V ( \cdot) $ is the variance, and $ C (\cdot,\cdot) $ is the covariance. A similar equation can be derived for $ s_{\text{f}} $. In the second line we have made a mean-field expansion (see Appendix). We have also assumed an independent distribution where $ p_{ij} = p_i p_j $ for this case. This is consistent with Hardy-Weinberg equilibrium. For cases with zero variance as the initial condition in (\ref{selectioneffective}), the time evolution is static.  However, any small but finite population with a higher viability genotype ($ A_N A_N $ in this case) will eventually dominate the population, occurring at a timescale (\ref{timeselection}). The population in $ A_N A_N $ will overtake $ A_1 A_1 $ in frequency due to its higher viability.  However, this needs some small seed population to instigate the process.

\subsection{Mutation and critical wait time}

The effect of mutation is to randomly distribute the genotypes. It is particularly relevant in the scenario with two competing genotypes considered, where the entire population is initially present in the less fitter of the two ($A_1 A_1 $ in Fig. \ref{fig3}(a)).  Assuming a small mutation rate $ v_{ij} \ll \gamma_{ij} $, a proportion of the population eventually always transitions to the genotype with the higher viability.  This is consistent with (\ref{timeselection}) where the time required for this to occur is dependent on the difference in mortality rate.  Interestingly, we observe a critical wait time for very small mutation rates before this transition starts to take place, as can be observed from Fig. \ref{fig3}(c)(d)(e). 

We now derive the time required before the change in population starts to occur in the presence of mutation. Taking the scenario in Fig. \ref{fig3}(c)(d) where we have two dominant genotypes with high viability, as shown in Fig. \ref{fig3}(a), we approximate the probability distribution as arising from two main contributions
\begin{align}
\frac{d p_{11}}{dt}  = & - \gamma_{11} p_{11} + (\gamma_{11} p_{11} + \gamma_{NN} p_{NN}) p_{11} \nonumber \\
& - 2 v p_{11} +v (p_{1N} + p_{N1} ) + r p_{11}^2 - r p_{11}   \nonumber \\
\frac{d p_{NN}}{dt} = &  - \gamma_{NN} p_{NN} + (\gamma_{11} p_{11} + \gamma_{NN} p_{NN}) p_{NN} \nonumber \\
& - 2 v p_{NN} +v (p_{1N} + p_{N1} )+ r p_{NN}^2 - r p_{NN}  .
\end{align}
Due to the low viability of $ p_{1N} $ and $ p_{N1} $, and population that mutates into these genotypes do not survive, and are redistributed equally to  $ p_{11} $ and $ p_{NN} $. Thus we can set $ p_{1N} = p_{N1} = (p_{11} + p_{NN})/2 $.  Defining the variable $ s = p_{11} - p_{NN} $, we have
\begin{align}
\frac{ds}{dt} = \Delta \gamma (s^2 - 1) - 2 v s .
\end{align}
This can be solved analytically with solutions
\begin{align}
s = \frac{v}{\Delta \gamma} - \frac{\sqrt{v^2 + \Delta \gamma^2}}{\Delta \gamma} \tanh ( \sqrt{v^2 + \Delta \gamma^2} (t - t_{\text{c}} ))
\end{align}
If the whole population is at $ p_{11} $ initially, then at $ t = 0 $, $ s = 1 $, and we have
\begin{align}
t_{\text{c}} &  = \frac{\tanh^{-1}( \frac{\Delta \gamma - v}{ \sqrt{v^2 + \Delta \gamma^2}})}{\sqrt{v^2 + \Delta \gamma^2}} \nonumber \\
& \approx \frac{\ln (\Delta \gamma/v)}{\Delta \gamma}
\label{criticaltime}
\end{align}
where for large $ x $ we have approximated $ \tanh(x) \approx 1 - 2 e^{-2x} $ and assumed that $ v \ll \Delta \gamma $.  
The critical wait time for the parameters in Fig. \ref{fig3}(e) correspond to  $ t_{\text{c}}  = 23, 46, 69 $ for $ v = 10^{-2}, 10^{-3}, 10^{-4} $ respectively. This agrees well with the numerics. The region of the crossover can be identified by a high genetic diversity in the population distribution (Fig. \ref{fig3}(g)).

Given that $ t = 0 $ precedes the appearance of a mutation, the critical wait time $ t_{\text{c}} $ can be attributed to the process and likelihood of obtaining a viable transmittable mutation in an individual being naturally small.  However, once the critical time (\ref{criticaltime}) is exceeded, the population's transition to the superior genotype can occur quickly on a timescale of (\ref{timeselection}) due to the propagation of the mutation through the reproductive process leading to a lower death rate as a result of an enhanced viability. The logarithmic dependence of (\ref{criticaltime}) also shows that even for extremely small mutation rates (i.e. exponentially small) the critical time occurs at a relatively short timescale.

\subsection{Gene flow and genetic drift}

 In order to simulate gene flow, we consider the situation where a small group of individuals with a genotype that has a higher viability than the rest of the population is introduced. We observe that the phenomenon of gene flow overcomes the critical wait time.  In Fig. \ref{fig3}(f) a small seed population in the most viable $ A_N A_N $ genotype is introduced at $ t = 0 $.  We see that for all mutation rates the population immediately shifts towards the more viable genotype. This is essentially independent of the mutation rate since the critical population is already introduced as the initial condition thereby overriding the process and critical time required for a viable mutation to emerge in a population.  This shows that the effect of introducing mutation is not strictly equivalent to introducing a seed population since  there is no critical wait time in the latter case.  Another interesting feature of the critical wait time is that it is robust even under initial conditions that do not perfectly fall in the less viable genotype.  For our results in Fig. \ref{fig3}(c)(d)(e)(g) we start with a distribution that mixes genotypes in the vicinity of $ i = j = 1$.  The distribution initially relaxes completely into  $ i = j = 1$ before making the transition to the more viable genotype  $ i = j = N$.

In the above examples, we have not discussed explicitly how genetic drift can occur within the framework.  In a probabilistic framework, opportunity for genetic drift occurs when the variance of the distribution is large.  A large variance means that genetic diversity is allowable, as shown in Fig. \ref{fig3}(g)(h).  This amounts to the possibility of the dominant genotype not being fixed and drifting in time.  To see this explicitly one would perform a stochastic simulation of (\ref{master}) and observe the genetic distribution.  At times when the variance of the distribution is small (such as the start and end points of Fig. \ref{fig3}(g)), little genetic drift is possible because the dominant genotype is more widely distributed.

\section{Entropy production and Stability of the system}
\label{sec:entropy}

Up to this point, our analysis of the master equation (19) has been using observables such as probabilities, expectation values, and variances illustrating the changes in population distribution. This is the conventional approach taken in numerous population genetics studies \cite{Gil04}.  However, we may equally take the point of view that it is a statistical system governed by a master equation, which can analyzed using techniques derived from statistical mechanics  \cite{schnakenberg1976network}.  The time dependent evolution towards steady-state that we have examined is then a non-equilibrium problem, hence we must use concepts derived from non-equilibrium statistical mechanics.  In this section, we illustrate this point by studying the entropy production and stability of the system.    

A thermodynamic system which is not in equilibrium is exposed to a set of external 
perturbations or driving forces.  These thermodynamic forces result in establishing a set of fluxes which
move the system from one state to another.  From the genotype probabilities $p_{\bm{i} \bm{j}}$ we can calculate the entropy using the standard expression for the entropy
\begin{align}
S = - \sum_{\bm{i}\bm{j}} p_{\bm{i}\bm{j}} \ln p_{\bm{i}\bm{j}} .
\label{entropydef}
\end{align}
Here we would like to note that this is only the {\it Shannon} (i.e. informational) entropy of the system.  
In a population genetics context the thermodynamic entropy far outweighs the contribution of the 
informational entropy due to the physical manifestations of the organisms.  But the informational 
entropy can still give us a characterization of the distribution of the population at any given time. 
For a completely homogeneous population such us that given in Fig. \ref{fig2}(a), the entropy is 
$ S = 0 $, while for a highly diverse population the entropy is large.

\begin{figure}
\includegraphics[width=\columnwidth]{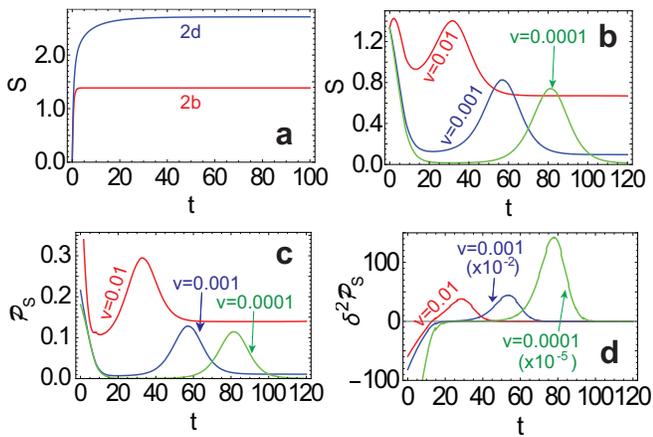}
\caption{	\label{fig4} Entropy and entropy production during the evolutionary process.  (a) The entropy (\ref{entropydef}) for the processes in Fig. \ref{fig2}(b)(d).  Lines are marked according to the figure numbers.   (b) The entropy (\ref{entropydef}) for the processes in Fig. \ref{fig3}(e)(g).  Lines are marked according to the mutation rate. (c) The total entropy production (\ref{entropyprod}) for the process in Fig. \ref{fig3}(e)(g). (d) The Glansdorff-Prigogine stability of the steady state (\ref{gplcondition}) is shown using a graph of the second variation of the entropy production as a function of time for different viabilities. The steady-state probabilities are approximated by the probabilities at the end of the time evolution $ \bar{p}_{\bm{i}\bm{j}} \approx p_{\bm{i}\bm{j}} (t_{\text{max}}) $.  The curves for $v= 0.001$ and $v=0.0001$ have been multiplied by the factors as labeled to fit in the plots.  }
\end{figure}

The entropy $ S $ can either increase or decrease depending on the particular 
dynamics of the master equation.  This can be seen from Fig. \ref{fig4}(a)(b), which shows the 
entropy (\ref{entropydef}) in Figs. \ref{fig2}(b)(d) and \ref{fig3}(e)(g) respectively.  In the 
case of Fig. \ref{fig4}(a), the dynamics only consists of an initially homogenous population 
(with a heterozygous genotype) evolving into a more diverse population.  This is accompanied 
by a consistent increase in the Shannon entropy in the system, as expected.  Meanwhile, 
Fig. \ref{fig4}(b) shows a more complex behavior, where the entropy initially decreases, then 
is followed by a time period with large entropy, then again settling down to a lower 
steady-state entropy.  This can be understood according to the gene flow dynamics as described 
earlier. The initial decrease of entropy occurs due to the initial state being distributed in 
the region of the 2nd most viable genotype $ i = 1 $, $ j = 1 $ becoming more concentrated in 
this corner, as also evident from Fig. \ref{fig3}(e).  Depending on the mutation rate $ v $, 
gene flow then occurs at a later time, which is accompanied by a period of high entropy.  
This is consistent with the variances as shown in Fig. \ref{fig3}(g). 

A non-equilibrium system is generally accompanied by a production of entropy as a result of its
dynamics and coupling to a reservoir \cite{schnakenberg1976network}.  The entropy production is 
not merely the contribution due to the increase of entropy $ \frac{dS}{dt} $, but also has a contribution
due to an external set of thermodynamic forces from the reservoir.  We follow the methods of 
Ref. \cite{schnakenberg1976network} to estimate the macroscopic entropy production for our system.  
The thermodynamic forces can be defined by first identifying the thermodynamic flux, which is its 
conjugate quantity.  We first write the master equation in its general form
\begin{align}
\frac{d p_{\bm{i} \bm{j}}}{dt} = & \sum_{\bm{k} \bm{l}} J_{\bm{i} \bm{j};\bm{k} \bm{l}} =  \sum_{\bm{k} \bm{l}} \left[  J^+_{\bm{i} \bm{j};\bm{k} \bm{l}} -  J^-_{\bm{i} \bm{j};\bm{k} \bm{l}} \right],
\label{currentmaster}
\end{align}
where the $  J^{\pm}_{\bm{i} \bm{j};\bm{k} \bm{l}} $ are the positive and the negative probability currents that contribute to the 
gain or loss of the probability $p_{\bm{i} \bm{j}}$.  The explicit expressions for the master equation (\ref{master}) is given in the Appendix. The associated thermodynamic force is then
\begin{align}
\mathcal{F}_{\bm{i} \bm{j};\bm{k} \bm{l}} = \ln \frac{J^+_{\bm{i}\bm{j};\bm{k}\bm{l}}}{J^-_{\bm{i}\bm{j};\bm{k}\bm{l}}}.
\label{force}
\end{align}
From the above expressions, we can write the bilinear expression which 
gives the entropy production in the system \cite{schnakenberg1976network}.  Thus the entropy production in the system is 
\begin{align}
{\cal P}_S = \sum_{\bm{i}\bm{j}\bm{k}\bm{l}} (  J^{+}_{\bm{i}\bm{j};\bm{k}\bm{l}} - J^{-}_{\bm{i}\bm{j};\bm{k}\bm{l}} ) \ln \frac{ J^{+}_{\bm{i}\bm{j};\bm{k}\bm{l}}}{ J^{-}_{\bm{i}\bm{j};\bm{k}\bm{l}}} .
\label{entropyprod}
\end{align}
This is a positive quantity and 
thus guarantees that the entropy production is always positive, i.e. the second law of thermodynamics is not violated.  This is in contrast to the entropy change $ \frac{dS}{dt} $ which can be negative, as can be observed from Fig. \ref{fig4}(b).  

Figure \ref{fig4}(c) show the macroscopic entropy production for the same process as that given in Fig. \ref{fig3}(e). We see that the entropy production is large during times where there is a population migration in the system.  Initially, the there is a population migration as can be observed from Fig. \ref{fig3}(e) since the initial state has broadening around the second most viable state.  Another peak is observed during the gene flow stage when the population transitions to the most viable genotype.  The positions of the peaks correspond to the times when the transitions occur, as can be seen in Fig. \ref{fig3}(e).  This is 
consistent with the findings in Refs. \cite{perunov2016statistical,england2015dissipative}, where periods of entropy production in the bath and drift towards a more likely outcome are different aspects of the same fundamental process. 

The stability of a non-equilibrium steady state can be evaluated from the Glansdorff-Prigogine stability criterion \cite{glansdorff1970non,glansdorff1974thermodynamic}.  
In terms of the probabilities this relation can be expressed as:
\begin{align}
\delta^{2} {\cal P}_S  = \frac{d}{dt} \left[ - \sum_{\bm{i}\bm{j}} 
\frac{1}{\bar{p}_{\bm{i}\bm{j}}} ( p_{\bm{i}\bm{j}} - \bar{p}_{\bm{i}\bm{j}} )^{2}
                                                   \right] \geq 0,
\label{gplcondition}                                                   
\end{align}
where ${\bar{p}_{\bm{i}\bm{j}}}$ is the steady state probability of the system. The above criterion can be understood to be a  manifestation of the Lyapunov stability criterion.  In our model
\begin{align}
\delta^{2} L  = -\sum_{\bm{i}\bm{j}}  \frac{1}{\bar{p}_{\bm{i}\bm{j}}} ( p_{\bm{i}\bm{j}} - \bar{p}_{\bm{i}\bm{j}})^{2}
\label{Lyapunov}
\end{align}
is the corresponding Lyapunov function.  In Fig. \ref{fig4}(d) we show the stability function (\ref{gplcondition}) 
which should be positive for stability. We see that in the vicinity of steady-state the criterion is always positive, 
indicating stability.  The initial negative values arise because of the particular dynamics that are present in the 
gene flow.  Due to the broadened population distribution chosen initially, the population initially migrates to the
2nd most viable genotype as discussed above.  This corresponds to a movement in the opposite direction to the eventual 
steady-state genotype in the opposite corner of Fig. \ref{fig3}(a).  Stability is define generally only in the vicinity 
of steady-state, hence this initial transient behavior has not reflection on the stability of the system.  While the above is only one example of the stability in the system, we expect that the dynamics should always give stable behavior for any choice of static parameters.   Investigations
for a driven system (i.e. time varying parameters) may on the other hand possibly show a different pattern of emergence of the stability.

\section{Summary and conclusions}

\label{sec:conc}

We have derived an explicit probabilistic master equation (\ref{master}) in the genotype space that incorporates selection, mutation, mating, and recombination for the evolutionary process.  Phenomena such as gene flow are consequences of the model, and are well-observed even for the simple case studies that we examine.   
While we only considered some simple toy models to illustrate the pertinent aspects of the model, 
we see no reason why this could not be extended to more realistic, sophisticated systems.  
We have shown that despite the rather simple models examined, the interplay of the various processes can produce interesting effects.  Using a prototypical fitness function model of one and two loci with $N$ possible allele variations existing in the population, an analytic formula for the time required to reach steady-state can be predicted. For the case including mutation, we find that there is a critical wait time before the genotype with the highest viability is found.  While we have only considered the diploid case in this paper, it is straightforward to generalize this to the polyploid case \cite{De71}. 

The formulation as a master equation of standard form allows 
for the application of this system to the tools of non-equilibrium statistical mechanics.
We have illustrated this by calculating the Shannon entropy production for the larger system including the bath.  This was found to have a behavior consistent with recent results which state that entropy production is largest during periods of migration towards more favorable states \cite{perunov2016statistical}. Finally, we evaluated the stability of the system using the Glansdorff-Prigogine criterion, and found that the non-equilibrium steady-state is stable.  These examples are merely illustrative of how results from non-equilibrium statistical mechanics can be carried over to systems in evolutionary biology. Calculation of the entropy and stability is convenient from the point of view that these can be calculated using only the probability distribution.  However, it should be possible to define and calculate other macroscopic thermodynamic quantities characterizing the system. The entropy production should also be relatable to Crook's microscopic reversibility relation \cite{crooks1998nonequilibrium}.  Treating the model as a non-equilibrium statistical model is interesting not only from the point of development of new techniques to treat such systems, but also in the investigations of the foundations of life \cite{england2015dissipative}. We note that the entropy production that we discuss in this article is the {\it informational} entropy and not the {\it thermodynamic} entropy, since our master equation is in the genotype space.   In this paper we only take into account the 
genetic information during evolution, but in fact in real biological system, there will be 
in addition the entropy contribution due to the biological machinery.  Nevertheless, this contributes to the total entropy of the system, and can be used as a characterization tool to understand the evolutionary process.  

The master equation we have introduced has interesting parallels with other non-equilibrium systems in statistical mechanics. For instance, the situation as depicted in Fig. \ref{fig1}(a) can be compared to spin glass models where various spin configurations have different energies. This is in agreement with past works which have shown equivalences between models of genetics and spin models \cite{baake1997ising,hermisson2001four,saakian2004solvable,saakian2008evolution}.  Mutation in this case is analogous to thermal noise in spin glass systems. However, unlike spin glasses, which are a closed system in terms of the number of spins, here the system is open through the reproduction and death processes.  This adds another layer of complexity to the dynamics. Another difference to spin glasses is in the way that mutations can occur between various genotypes.  For this paper, we assumed a mutation matrix $v_{\bm{i}\bm{j}} $ where any allele can mutate into any other allele.  More realistically, mutations are more likely to occur as a result of a change in a single or a small number of nucleotides thus leading to alleles with similar genetic sequences  \cite{sanjuan04}, giving rise to a more complex mutation matrix. Similarly, certain spin configurations can only mutate into a few other spin configurations. As is well-known, finding the ground state of a spin glass is  equivalent to computational optimization problems.  An interesting question is then whether the open nature of the evolutionary problem affects the complexity of finding the most viable state.  Such problems, also encountered in the context of machine learning and quantum adiabatic computing, are of fundamental interest to a variety of fields.

\begin{acknowledgments}
We thank Jeremy England for discussions and Kourosh Salehi-Ashtiani for valuable comments regarding the manuscript. This work is supported by the National Natural Science Foundation of China (Grant No. 61571301); the Thousand Talents Program for Distinguished Young Scholars (Grant No. D1210036A); and the NSFC Research Fund for International Young Scientists (Grant No. 11650110425); the Science and Technology Commission of Shanghai Municipality (Grant No. 17ZR1443600); and the China Science and Technology Exchange Center (NGA-16-001). 
\end{acknowledgments}

\appendix

\section{Mean field approximation}

On the right hand side of  (\ref{selectioneffective0}), we have expectation values of second and third powers of $ s_{\text{m,f}}$ (moments), which in turn require evolution equations of themselves.  To obtain a closed set of equations, we approximate such moments up to a fixed order, by performing a mean field approximation.  For example, the covariance $ C(s^2,s) =  E(s^3) - E(s^2) E(s) $ involves a third power expectation value of $ s $, where $ E(s) = \bar{s} $. The mean field approximation is performed by first rewriting the random variable as $ s = E(s) + (s - E(s))  $ and taking the term in the brackets to be small. The third order expectation value is then
\begin{align}
E(s^3)   =& E\{ [E(s) + (s - E(s))]^2 [E(s) + (s - E(s))] \} \nonumber \\
 =& E \{  [ E(s)^2 + 2 E(s) (s - E(s))  + (s - E(s))^2 ] \nonumber \\
& \times [E(s) + (s - E(s))] \} \nonumber \\
 \approx &E(s)^3 + 3 E(s) E( (s -  E(s))^2 ) \nonumber \\
 = & 3  E(s^2)  E(s) - 2  E(s)^3
\end{align}
where in the second last line we dropped the term $ (s - E(s))^3 $. The covariance can then be approximated
\begin{align}
C(s^2,s) \approx 2 E(s) V(s) ,
\end{align}
which is accurate as long as the probability distribution assumes a form that can be approximated to a Gaussian in the variables $ s_{\text{m,f}}$.  The mean-field approximation is most valid in the limit where the number of alleles $N $ is large.

\section{Total entropy production}

Starting with the master equation (\ref{master}), we can introduce factors of $ \sum_{\bm{k} \bm{l}} p_{\bm{k} \bm{l}} = 1 $ and Kronecker deltas to make explicit the transitions between genotype $ A_{\bm{i}} A_{\bm{j}} $ and $ A_{\bm{k}} A_{\bm{l}} $
\begin{align}
\frac{d p_{\bm{i} \bm{j}}}{dt} = &  \sum_{\bm{k} \bm{l}}  \Big[ - \gamma_{\bm{i} \bm{j}} p_{\bm{i} \bm{j}} p_{\bm{k} \bm{l}} 
+ \gamma_{\bm{k} \bm{l}} p_{\bm{k} \bm{l}}  p_{\bm{i} \bm{j}} \nonumber \\
& - ({v}_{\bm{j} \bm{k}} \delta_{\bm{i} \bm{l}} + {v}_{\bm{i} \bm{k}} \delta_{\bm{j} \bm{l}} ) p_{\bm{i} \bm{j}} +  ( v_{\bm{k} \bm{j}} \delta_{\bm{l} \bm{i}}  p_{\bm{l} \bm{k}} +  v_{\bm{k} \bm{i}} \delta_{\bm{l} \bm{j}} p_{\bm{k} \bm{l}}) \nonumber \\
& + \frac{r}{4} \delta_{\bm{i} \bm{l}}  \left( p_{\bm{l} \bm{k}}  + p_{\bm{k} \bm{l}} \right) \sum_{\bm{k}' \bm{l}'} 
\delta_{\bm{j} \bm{l}'}   \left( p_{\bm{l}' \bm{k}'} +  p_{\bm{k}' \bm{l}' } \right) - r p_{\bm{i} \bm{j}} p_{\bm{k} \bm{l}} \Big],
\end{align}
where we have set the recombinant probability $ c_{nm} = 0 $ for simplicity. Noting that all coefficients are positive, we can interpret positive terms as being gain terms from the $ A_{\bm{k}} A_{\bm{l}} $ to $ A_{\bm{i}} A_{\bm{j}} $, and negative terms as loss terms due to transitions between $ A_{\bm{i}} A_{\bm{j}} $ to $ A_{\bm{k}} A_{\bm{l}} $.  The master equation then can be written as (\ref{currentmaster}), where the positive and negative currents are 
\begin{align}
J^+_{\bm{i} \bm{j};\bm{k} \bm{l}} & = \Big[ \gamma_{\bm{k} \bm{l}} p_{\bm{i} \bm{j}} + v_{\bm{l} \bm{j}} \delta_{\bm{k} \bm{i}} +  v_{\bm{k} \bm{i}} \delta_{\bm{l} \bm{j}}  \nonumber \\
& +  \frac{r}{8} ( \delta_{\bm{i} \bm{k}} + \delta_{\bm{i} \bm{l}})
 \sum_{\bm{k}'}  \left( p_{\bm{j} \bm{k}'} +  p_{\bm{k}' \bm{j} } \right) \nonumber \\
& +  \frac{r}{8} ( \delta_{\bm{j} \bm{k}} + \delta_{\bm{j} \bm{l}})
 \sum_{\bm{k}'}  \left( p_{\bm{i} \bm{k}'} +  p_{\bm{k}' \bm{i} } \right)
 \Big] p_{\bm{k} \bm{l}} 
\end{align}
and
\begin{align}
J^-_{\bm{i} \bm{j};\bm{k} \bm{l}} = \Big[ \gamma_{\bm{i} \bm{j}} p_{\bm{k} \bm{l}} 
+ v_{\bm{j} \bm{k}} \delta_{\bm{i} \bm{l}} + v_{\bm{i} \bm{k}} \delta_{\bm{j} \bm{l}} 
+ r  p_{\bm{k} \bm{l}}  \Big]  p_{\bm{i} \bm{j}} .
\end{align}
%

% How to do the references:
%% 1) First uncomment the below and compile
%\bibliographystyle{apsrev}
%\bibliographystyle{naturemag}
%\bibliography{references}
%% 2) Copy the .bbl file to below and comment out the above two lines.

\end{document}